\documentclass[useASM, referee]{mn2e}
\input{psfig.sty}
\usepackage{graphicx}
\begin{document}
\title{Spectral lags caused by the curvature effect of fireballs}
\date{2005 February 20}
\pubyear{????} \volume{????} \pagerange{2} \onecolumn
\author[Lu et al. ]
       {R.-J.\ Lu$^{1,2,3,4}$, Y.-P. \ Qin $^{1,2}$, Z.-B. \ Zhang $^{1,3}$, T.-F.\ Yi$^{2}$ \\
$^1$National Astronomical Observatories/Yunnan Observatory,Chinese Academy of Sciences,\\
P. O. Box 110, Kunming 650011, China\\
$^2$Physics Department, Guangxi University, Nanning, Guangxi 530004, P. R. China\\
$^3$The Graduate School of the Chinese Academy of Sciences; \\
$^4$E-mail:luruijing@126.com
 }
\pagestyle{plain}
\date{Accepted ????.
      Received ????;
      in original form 2005 June 10}
\pagerange{\pageref{firstpage}--\pageref{lastpage}} \pubyear{2004}
\maketitle \label{firstpage}
\begin{abstract}
Recently, Shen et al. (2005) studied the contributions of curvature
effect of fireballs to the spectral lag and showed that the observed
lags could be accounted for by the effect. Here we check their
results by performing a more precise calculation with both the
formulas presented in Shen et al. (2005) and Qin et al. (2004).
Several other aspects which were not considered in Shen et al.
(2005) are investigated. We find that in the case of
ultra-relativistic motions, both formulas are identical as long as
the whole fireball surface is concerned. In our analysis, the
previous conclusion that the detected spectral lags could be
accounted for by the curvature effect is confirmed, while the
conclusion that the lag has no dependence on the radius of fireballs
is not true. We find that introducing extreme physical parameters is
not the only outlet to explain those observed large lags. Even for
the larger lags ($\sim5s$) (see Norris et al. 2005), a wider local
pulse ($\Delta t_{\theta,FWHM} = 10^{7}s$) could account for it.
Some conclusions not presented in Shen et al. (2005) or those
modified in our analysis are listed below: a) lag $\propto
\Gamma^{-\epsilon}$ with $\epsilon>$ 2; b) lag is proportional to
the local pulse width and the $FWHM$ of the observed light curves;
c) a large lag requires a large $\alpha_0$ and a small $\beta_0$ as
well as a large $E_{0p}$; d) when the rest frame spectrum varies
with time, the lag would become larger; e) lag decreases with the
increasing of $ R_c$; f) lag $\propto$ E within the certain energy
range for a given Lorentz factor; g) lag is proportional to the
opening angle of uniform jets when $\theta_j < 0.6\Gamma^{-1}$.
\end{abstract}


\section{Introduction}

It is common that there is a spectral lag between different energy
channels in gamma-ray bursts (see, e.g. Cheng et al. 1995; Norris et
al. 1996, 2000; Wu \& Fenimore 2000; Chen et al. 2005). Although
There are several attempts to explain the origin of the spectral lag
(See, e.g., Salmonson 2000; Ioka \& Nakamura 2001; Schaefer 2004;
Kocevski et al. 2003a), the problem remain unresolved.

Recently, Shen et al. (2005) (hereafter Paper I) tentatively studied
the contribution of curvature effect of fireball to the lag, and the
resulting lags are very closed to the observed one. In fact, in
their paper introducing extreme physical parameters such as $\Gamma
< 50$ or $\alpha > -0.5$ is not the only outlet to explain those
observed large lags. A reasonable intrinsic pulse width $t_{d}^{'}$,
i.e., hundreds to thousands of seconds, which could produce the lag
of $0.1 - 1$ s based on their model, could also account for it.

Qin (2002) had derived in a much detail the flux intensity based on
the model of highly symmetric expanding fireballs, where the Doppler
effect of the expanding fireball surface is the key factor to be
concerned, and no terms are omitted in his derivation, therefore the
formula is applicable to the cases of relativistic,
sub-relativistic, and non-relativistic motions and to those of
spherical fireballs or uniform jets. With this formula, Qin (2003,
2004) derived the observed photon count rate of a fireball (See Eq.
(1) in the paper).

A detailed comparison between Shen et al. (2005) and Qin et al.
(2004) model (hereafter Paper II) gives rise to the conclusion that
the two models are identical when the following conditions are
satisfied. Firstly, as shown in equation (8) of Paper II, the two
quantities, local time $t_{\theta}$ and observed time t, are related
by $t=(1-\beta cos \theta)(t_{\theta}-t_c)+t_c+D/c-R_c cos\theta/c$,
where $t_{c}$ is the initial local time, $R_{c}$ is the radius of
the fireball measured at $t_{\theta}=t_c$, and D is the distance of
the fireball to the observer. When taking $t_c=0$ and
$T=t+(R_c-D)/c$, one would find that the above relation will become
the one determined by equation (3) in Paper I (where T is referred
to as observed time defined in Paper I). Secondly, in equation (3)
of Paper II, the integral lower and upper limit are determined by
$\widetilde {\theta}_{\min}=cos^{-1} \min \{ cos \theta_{\min},
(t_{\theta, \max} - t + D/c) / ((t_{\theta, \max} - t_c)\beta +
R_c/c)\}$ and $\widetilde {\theta}_{\max}=cos^{-1} \max \{ cos
\theta_{\max}, (t_{\theta, \min} - t + D/c) / ((t_{\theta, \min} -
t_c)\beta + R_c/c)\}$, respectively, where $\theta_{\min}$ and
$\theta_{\max}$ are confined by the concerned area of the fireball
surface, and $t_{\theta, \min}$ and $t_{\theta, \max}$ are confined
by the local time range of emission. For such a local pulse,
$t_{\theta, \min}=0$ and $t_{\theta, \max} \rightarrow \infty $,
when taking $\theta_{\min}=0$ and $\theta_{\max}=\pi/2$ (i.e., the
photons reaching the observer come from the whole surface of the
fireball), one would obtain $\widetilde {\theta}_{\min}=0 $ and $
\widetilde {\theta}_{\max}=cos^{-1}(1-T/\tau) $ (where
$\tau=R_c/c$), which are consistent with the integral lower and
upper limit in equation (4) of Paper I, respectively. In a word, the
mainly difference between the two models is that, Shen et al.
neglected the constraint of the local pulse and the concerned area
of fireball surface on the integral lower and upper limit of the
observed flux (see equation (4) in their paper), which plays a
important role in producing the observed light curves. Strictly,
when dealing with uniform jets with a small opening angle such as
$\theta<1/\Gamma$, the formula presented in Shen et al. (2005) would
fail to be applied.

We in this paper check their results by performing a more precise
calculation with the two models, at the same time, several other
aspects which were not considered in Shen et al. (2005) are
investigated, such as lag's dependence on energy and opening angle
(in Section 2). The previous conclusion that the observed spectral
lags could be accounted for by the curvature effect is confirmed and
some new results are obtained. For example, lag decreases with the
increase of radius of fireball other than weakly dependent on it.
Lag $\propto \Gamma^{-\epsilon}$, $\epsilon>2$ in the case of fixing
the local pulse width, which is identical with that, lag $\propto
\Gamma^{-1}$ in the case of fixing the intrinsic pulse width,
obtained by Shen et al. (2005). Accordingly an interesting question
arises: would the two approaches always lead to identical results?
which will be discussed in Section 3. And then we give our
discussions and conclusions in the last section.

\section{ Theoretical analysis on spectral lags }

Under the assumption that a fireball expands isotropically with a
constant Lorentz factor $\Gamma>1$ and the radiation is
independent of direction, the expected count rate of the fireball
measured within frequency interval $[\nu_1, \nu_2]$ can be
calculated with equation (21) in paper II, which can be rewritten
as follows
\begin{equation}
C(\tau )=\frac{2\pi R_c^3\int_{\widetilde{\tau }_{\theta ,\min }}^{%
\widetilde{\tau }_{\theta ,\max }}\widetilde{I}(\tau _\theta
)(1+\beta \tau
_\theta )^2(1-\tau +\tau _\theta )d\tau _\theta \int_{\nu _1}^{\nu _2}\frac{%
g_{0,\nu }(\nu _{0,\theta })}\nu d\nu }{hcD^2\Gamma ^3(1-\beta
)^2(1+\frac{\beta}{1-\beta}\tau )^2}.
\end{equation}
Note that the formula is in terms of the integral of
$\tau_{\theta}$, which is a dimensionless relative local time
defined by $ \tau _\theta \equiv c(t_\theta -t_c)/R_c$, where
$t_\theta$ is the emission time in the observer frame, called
local time, of photons emitted from the concerned differential
surface $ds_\theta $ of the fireball ($\theta$ is the angle to the
line of sight), $t_c$ is the initial local time which could be
assigned to any values of $t_\theta$, and $R_c$ is the radius of
the fireball measured at $t_\theta=t_c$. Variable $\tau$ is a
dimensionless relative observation time defined by $ \tau \equiv
[c(t-t_c)-D+R_c]/R_c$, where $D$ is the distance of the fireball
to the observer, and $t$ is the observation time measured by the
distant observer. In formula (1), $\widetilde{I}(\tau _\theta )$
represents the development of the intensity magnitude of radiation
in the observer frame, called as a local pulse function, and
$g_{0,\nu }(\nu _{0,\theta })$ describes the rest frame radiation
mechanisms.

Light curves determined by equation (1) are dependent on the
integral limits $\widetilde{\tau }_{\theta ,\min }$ and
$\widetilde{\tau }_{\theta ,\max }$, which are determined by the
concerned area of the fireball surface, together with the emission
ranges of the radiated frequency and the local time. But for
commonly adopted mechanisms, such as the bremsstrahlung,
Comptonized and synchrotron radiations, they cover the entire
frequency band, and do not provide constraints on the integral
limits (see equations (14) and (15) in Qin 2002). Thus the
integral limits are only determined by $ \widetilde{\tau }_{\theta
,\min }=\max \{\tau _{\theta ,\min },(\tau -1+\cos \theta _{\max
})/(1-\beta \cos \theta _{\max })\} $ and $ \widetilde{\tau
}_{\theta ,\max }=\min \{\tau _{\theta ,\max },(\tau -1+\cos
\theta _{\min })/(1-\beta \cos \theta _{\min })\},$ where
$\tau_{\theta, \min}$ and $\tau_{\theta, \max}$ are the lower and
upper limit of $\tau_{\theta}$ confining $\widetilde{I}(\tau
_\theta )$, and $\theta_{\min}$ and $\theta_{\max}$ are confined
by the concerned area of the fireball surface. The radiations are
observable within the range of $(1-\cos \theta _{\min })+(1-\beta
\cos \theta _{\min })\tau _{\theta ,\min } \leq \tau \leq (1-\cos
\theta _{\max })+(1-\beta \cos \theta _{\max })\tau _{\theta ,\max
}$ (see Paper II).

Because peak times of different light curves associated with
different frequency intervals $[\nu_1, \nu_2]$ or different energy
bands $[E_1, E_2]$ are different, following Shen et al. (2005), we
define the spectral lag as the time between the peaks of the light
curves in two different channels, a lower energy channel $[E_1,
E_2]$ and a higher energy channel $[E_3, E_4]$. In the following,
we will investigate on what parameters the lags are dependent in
terms of the fireball model.

For the sake of simplicity, we employ in the following a local
Gaussian pulse and employ the Band function as a rest frame
radiation spectrum which was frequently and rather successfully
employed to fit the spectra of GRBs (see, e.g., Schaefer et al.
1994; Ford et al. 1995; Preece et al. 1998, 2000) to study the
issue. A local Gaussian pulse and the Band function are written as
\begin{equation}
\widetilde{I}(\tau _\theta )=I_0\exp [-(\frac{\tau _\theta -\tau _{\theta ,0}%
}\sigma )^2]\qquad \qquad \qquad \qquad \qquad (\tau _{\theta ,\min
}\leq \tau _\theta ),
\end{equation}
and
\begin{equation}
g_{0,\nu }(\nu _{0,\theta })=\{
\begin{array}{c}
(\frac{\nu _{0,\theta }}{\nu _{0,p}})^{1+\alpha _0}\exp [-(2+\alpha _0)\frac{%
\nu _{0,\theta }}{\nu _{0,p}}]\quad (\frac{\nu _{0,\theta }}{\nu _{0,p}}<%
\frac{\alpha _0-\beta _0}{2+\alpha _0}) \\
(\frac{\alpha _0-\beta _0}{2+\alpha _0})^{\alpha _0-\beta _0}\exp
(\beta_0-\alpha _0)(\frac{\nu _{0,\theta }}{\nu _{0,p}})^{1+\beta _0}\quad (\frac{%
\nu _{0,\theta }}{\nu _{0,p}}\geq \frac{\alpha _0-\beta
_0}{2+\alpha_0})
\end{array},
\end{equation}
respectively, where $I_0$, $\sigma$ and $\tau_{\theta,min}$ are
constants, $\alpha_0$ and $\beta_0$ are the low and high energy
indices in the rest frame, respectively, and $\nu_{0,p}$ is the
rest frame peak frequency. Typical values of the low and high
energy indices coming from statistical analysis are $\alpha_0=-1$
and $\beta_0=-2.25$, respectively (see Preece et al. 1998, 2000).

Due to the constraint to the lower limit of $\tau_{\theta}$, which
is $\tau_{\theta,min}>-1/\beta$ (see Paper II), we assign
$\tau_{\theta,0} = 10\sigma+\tau_{\theta,min}$ so that the
interval between $\tau_{\theta,min}$ and $\tau_{\theta,0}$ would
be large enough to make the rising phase of the local pulse close
to that of the Gaussian pulse. From (2) one can obtain
$\Delta\tau_{\theta,FWHM} = 2\sqrt{ln 2}\sigma$, which leads to
$\sigma=\Delta\tau_{\theta,FWHM}/2\sqrt{ln 2}$, where
$\Delta\tau_{\theta,FWHM}$ is the FWHM of the Gaussian pulse. From
the relation between $\tau_{\theta}$ and $t_{\theta}$, one gets
$\Delta t_{\theta,FWHM}=(R_c/c)\Delta \tau_{\theta, FWHM}$. In the
following, we assign $(2\pi R_{c} ^{3} I_{0})/(hcD^{2})=1$,
$R_c=3\times10^{15}$ cm, $\tau_{\theta,min}= 0$, $\theta_{min}=0$,
and $\theta_{max}=\pi/2.$

\subsection{Lag's dependence on the Lorentz factor and the local pulse width }

As shown in Qin et al. (2004), the Lorentz factor and the local
pulse width are important in producing the light curve observed.
We wonder if the Lorentz factor and the local pulse width play
important roses in producing a lag. The spectral lag of GRBs could
be found either in hight BATSE energy channels or in lower Ginga
ones (see, e.g., Cheng et al. 1995; Chen et al. 2005; Norris et
al. 1996, 2000; Wu \& Fenimore 2000). We consider the following
three different energy channel pairs: BATSE channels 1 and 3,
BATSE channels 1 and 4, and Ginga channels between 2-10 keV and
50-100 keV (marked as $Lag_{13}$, $Lag_{14}$ and $Lag_{G}$,
respectively).

As suggested by observation, the value of the peak energy $E_{p}$
of the GRB's $\nu F_{ \nu }$ spectrum is mainly distributed within
$100 \sim 600$ keV (See Preece et al. 2000) and peaks at about 250
keV. Here we assume 250 keV and 50 keV as the typical values of
the peak energy of hard and soft bursts, respectively. In
addition, we assume that typical hard and soft bursts come from
the same Lorentz factor of the expanding motion of the fireball
surface and differ only in their rest fame peak energy $E_{0,p}$.
As derived in Qin (2002), when taking into account the Doppler
effect of fireballs, the observed peak energy would be related
with the peak energy of the typical rest frame Band function
spectrum ($\alpha_0=-1$, $\beta_0=-2.25$) by
$E_{p}\simeq1.67\Gamma E_{0,p}$. In a recent work, the Lorentz
factors of a GRB sample were found to be distributed mainly within
(100, 400) and to peak at about 200 (Qin et al. 2005a). So, when
assigning $\Gamma=200$ to be the Lorentz factor of these sources,
one would find the typical values of the rest frame peak energy
$E_{0,p}$ of hard and soft bursts are 0.75 keV and 0.15 keV
respectively.

Presented in Figs. 1 and 2 are the relationships between the lag
and the Lorentz factor $\Gamma$ as well as between the lag and the
local pulse width, $\Delta t_{\theta, FWHM}$.

We find from the left panel in Fig. 1 that each of the three kinds
of lag has a peak value. The Lorentz factor corresponding to this
peak value is denoted by $\Gamma_{p}$ (where subscript ``p" stands
for ``peak"). The lags increase with $\Gamma$ when $\Gamma<
\Gamma_{p}$; the contrary is observed when $\Gamma > \Gamma_{p}$.
For $E_{0,p}=0.75 keV$, the lags decrease with the Lorentz factor
in the law of $lag \propto \Gamma^{-\epsilon}$ with $\epsilon>$ 2
for the three kinds of lag. The figure exhibits $Lag_{G} <
Lag_{13} < Lag_{14}$ within the range of $\Gamma>100$ (note that
current GRB models suggest that the gamma-ray photons come from a
relativistically expanding fireball surface with Lorentz factor
$\Gamma>100$, see Piran 2005). The result and that obtained by
Shen et al. (2005) are identical (note that here we fix the local
pulse width instead of the co-moving pulse width). One can check
that, when the local pulse width is large enough, for a fixed rest
frame peak energy the observed peak energy would increase with
$\Gamma$ following the law of $E_{p} = 1.67 \Gamma E_{0,p}$ for
the adopted Band function spectrum (as shown in Qin 2002).
According to the law of $E_{p} = 1.67 \Gamma E_{0,p}$, the lag
will have a peak value when $E_{p}$ shifting from a higher energy
channel to a lower one, which could be observed in Fig. 1. Fig. 1
shows that the relationship between the lag and $\Gamma$ would
differ according to different rest frame peak energy $E_{0,p}$,
which suggests that, in producing a lag, not only the Lorentz
factor is important but also the rest frame peak energy plays a
role.

Fig. 2 shows that the lag increases with the local pulse width
following the law of $Lag \propto \Delta t_{\theta,FWHM}$, and it
obeys the law of $Lag_{G}<Lag_{13}<Lag_{14}$, the same as that
shown in Fig. 1. However, as suggested by Fig. 2, the dependence
of the lag on the local pulse width is much obvious than that on
the Lorentz factor $\Gamma$ and the rest frame peak energy
$E_{0,p}$ (see Fig. 1). In a recent work, Lu et al. (2005a) found
that $FWHM \propto\Delta t_{\theta,FWHM}$ based on the same model
(Qin 2002, Qin et al. 2004) , where $FWHM$ is the full width at
half-maximum of the GRB's light curve. Thus one could come to $lag
\propto FWHM$, which is in agreement with the observation that
wider pulses were found to produce longer lags (see Norris et al.
1996, 2001, 2005).

\begin{figure}
\centering
\includegraphics[width=5in,angle=0]{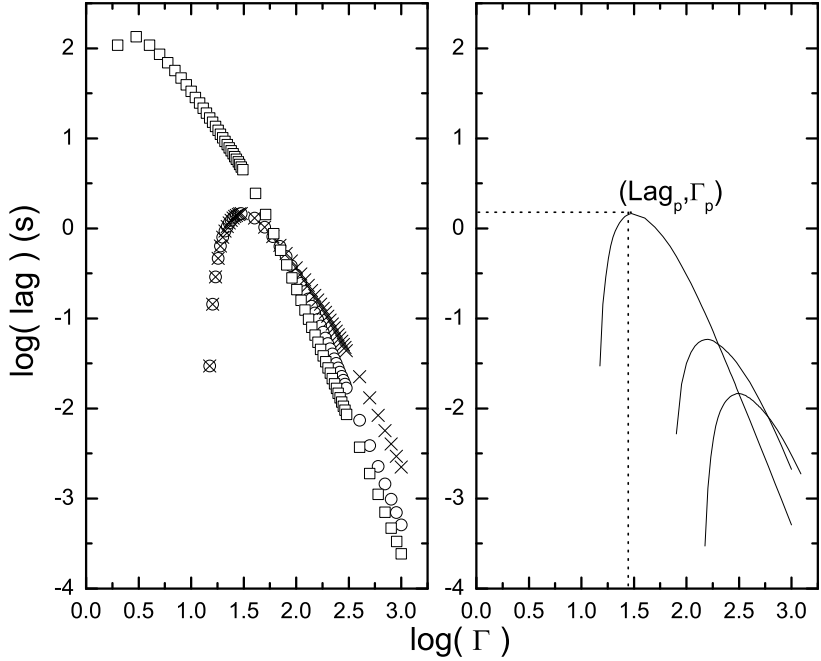}
\caption{--Relationships between $Lag_{13}$, $Lag_{14}$, $Lag_{G}$
and $\Gamma$ for $E_{0,p}$=0.75 keV (the left panel), and that
between $Lag_{13}$ and $\Gamma$ for different $E_{0,p}$ (the right
panel), where the Band function rest frame radiation form with
$\alpha_{0}=-1$ and $\beta_{0}=-2.25$ is adopted and we take
$R_c=3\times10^{15}$ cm and $ \Delta t_{\theta,FWHM}=10^{5} s$.
The open circle, the cross and the open square in the left panel
stand for $Lag_{13}$, $Lag_{14}$ and $Lag_{G}$, respectively. In
the right panel, solid lines from the top to the bottom stand for
$E_{0,p}$=0.75, 0.15, 0.075 keV, respectively.}
\end{figure}
\begin{figure}
\centering
\includegraphics[width=5in,angle=0]{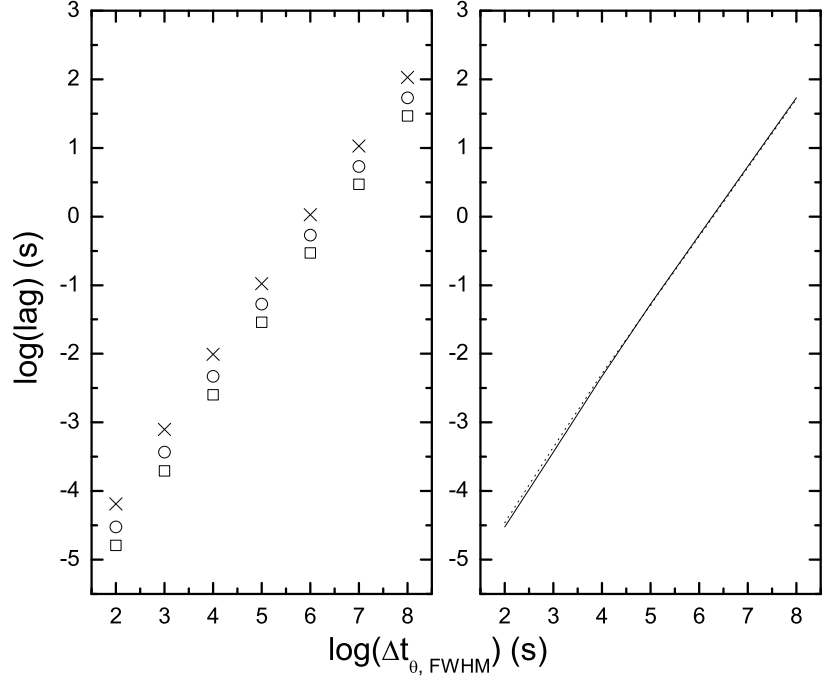} \caption{--Relationships between $Lag_{13}$, $Lag_{14}$,
$Lag_{G}$ and the local pulse width for  $E_{0,p}$=0.75 keV (the
left panel), and that between $Lag_{13}$ and local pulse width
(the right panel) for $E_{0,p}$=0.75 keV (the solid line) and 0.15
keV (the dotted line). Here we take $\Gamma=200$. Other parameters
and symbols are the same as those adopted in Fig. 1.}
\end{figure}

\subsection{Lag's dependence on spectral parameters of the Band function}

We know that different values of the spectral parameters (i.e.,
$\alpha_0$, $\beta_0$ and $E_{0,p}$) of the Band function are
associated with different mechanisms (see Band et al. 1993). Let
us study the impact of the mechanisms on the spectral lag in the
following.

We first investigate the lag's dependence on the observed peak
energy of the $\nu F_{\nu}$ spectrum. Ignoring the minute
difference caused by various factors, we assume that the following
relationship holds in any situations concerned in this paper:
$E_{p} = 1.67 \Gamma E_{0,p}$ (one can check that, for photons
emitted from an ejecta moving towards the observer, $E_{p} = 2
\Gamma E_{0,p}$). This relationship tells us that the observed
peak energy is proportional to $\Gamma$ or $E_{0,p}$ when the
other is fixed. The lag's dependence on the observed peak energy
in the case of $E_{p}$ varying with $\Gamma$ when fixing $E_{0,p}$
is similar to those discussed in $\S 2.1$. For the sake of
comparison, we convert the right panel of Fig. 1 to the last panel
of Fig. 3 by applying the relationship. In addition, we explore
the lag's dependence on the observed peak energy in the case of
$E_{p}$ varying with $E_{0,p}$ when fixing $\Gamma$.

\begin{figure}
\centering
\includegraphics[width=5in,angle=0]{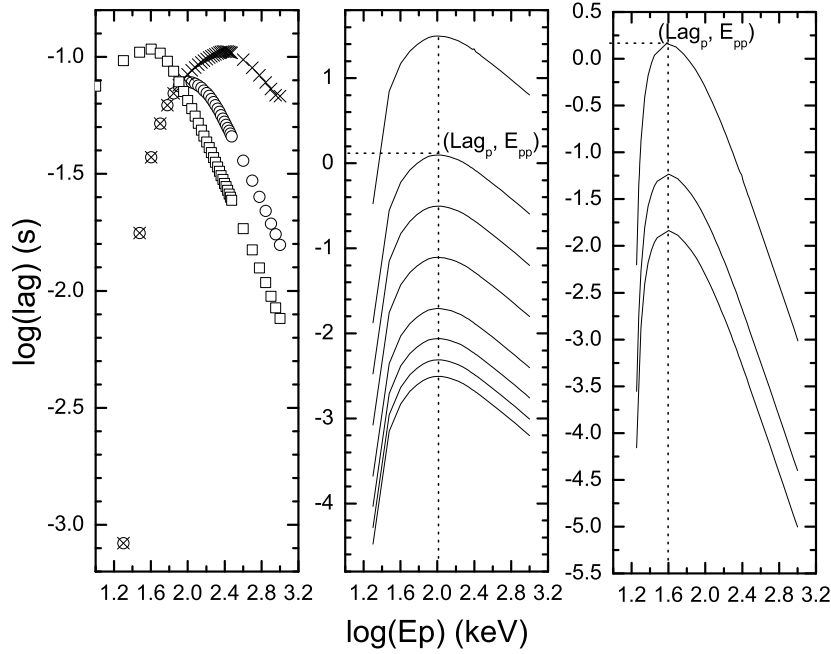} \caption{--Relationships between $Lag_{13}$, $Lag_{14}$,
$Lag_{G}$ and $E_p$ for $\Gamma=200$ (the left panel), and that
between $Lag_{13}$ and $E_p$ for the two cases of fixing $\Gamma$
(the middle panel) and $E_{0,p}$ (the last panel). In the middle
panel, the solid lines from the top to the bottom represent
$\Gamma$=10, 50, 100, 200, 400, 600, 800, 1000, respectively, and
in last panel, the solid lines from the top to the bottom stand
for $E_{0,p}$=0.75, 0.15, 0.075 keV, respectively. Other
parameters and symbols are the same as those adopted in Fig. 1.}
\end{figure}

One would find from the left panel of Fig. 3 that, similar to Fig.
1, when $E_{p}$ varies with $E_{0,p}$ for a fixed $\Gamma$, each
of the three kinds of lag has a peak lag ($Lag_{p}$) as well.

To make clear what causes the relationships shown in Fig. 3, we
divide the light curves of equation (1) into two parts, with one
being contributed from the low-energy portion of the rest frame
Band function spectrum and the other from the high-energy portion,
and explore how the peak times, $t_p$, of the light curves are
affected by the observed peak energy $E_{p}$ when it shifting
through the channel (see Figs. 4 and 5).

\begin{figure}
\centering
\includegraphics[width=5in,angle=0]{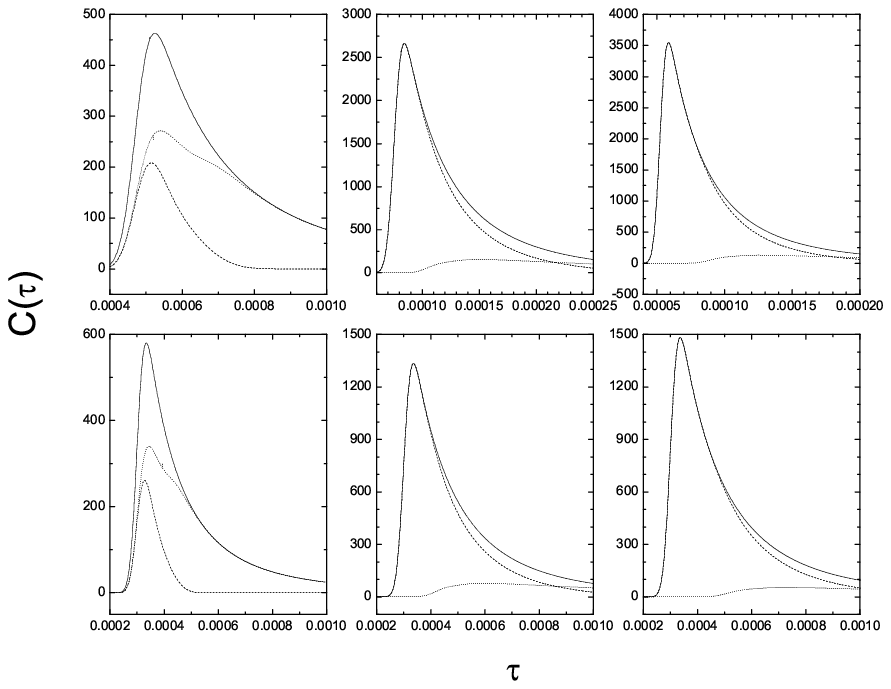} \caption{--Plots of the
light curves of BATSE channel 3 contributed from the rest frame
low-energy portion (the dash line) and high-energy portion (the
dot line), and from the whole energy range (the solid line) of a
Band function spectrum in cases of fixing $E_{0,p}=0.75 keV$ (the
upper panels) and $\Gamma=100$ (the lower panels). The panels from
the left to the right stand for $E_{p}$=100 keV, 250 keV, 300 keV,
respectively. Other parameters are the same as those adopted in
Fig. 1.}
\end{figure}

\begin{figure}
\centering
\includegraphics[width=5in,angle=0]{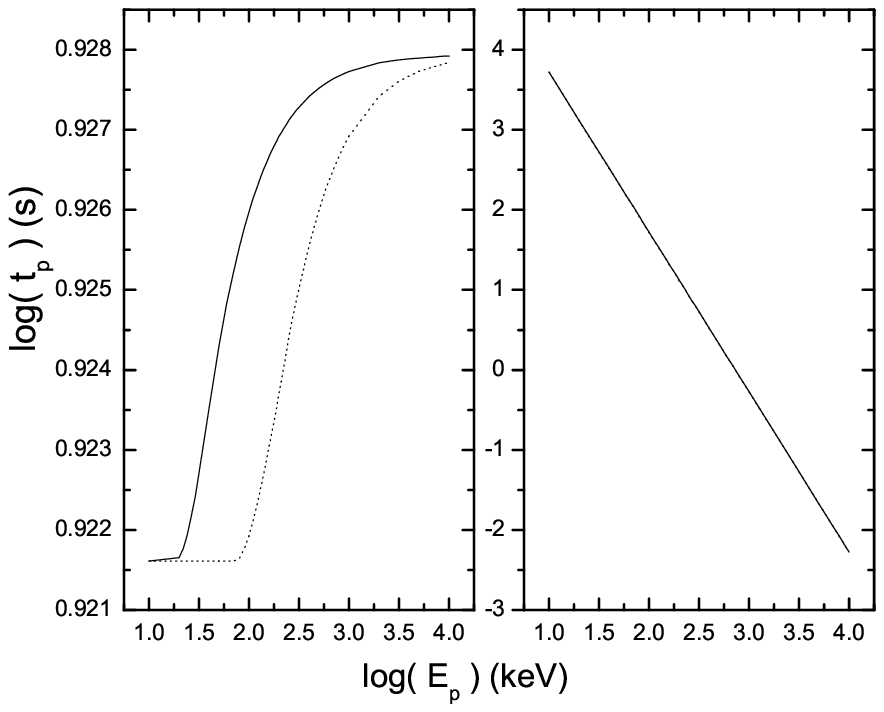} \caption{--Relationships between $E_{p}$
and the peak time $t_p$ of the light curves of BATSE channels 1 and
3 in cases of fixing $\Gamma=200$ (the left panel) and $E_{0,p}=0.75
keV$ (the right panel). Here we take $R_c-D-ct_c=0$. Other
parameters are the same as those adopted in Fig. 4. The solid and
the dot line stand for BATSE channels 1 and 3, respectively. (Note
that the two lines in the right panel almost overlap each other.)}
\end{figure}

Let us denote the channel concerned by $[E_{1}, E_{2}]$ (for the
first channel of BATSE, it is $[E_{1}, E_{2}]=[25,50]keV$, while
for the third channel, it is $[E_{1}, E_{2}]=[100,300]keV$). From
Figs. 4 and 5 some conclusions could be reached. For a certain
value of the Lorentz factor (here we take $\Gamma=100$), when
$E_{p}<E_1$, the contributions to the light curves mainly come
from the rest frame high-energy portion. According to the Doppler
effect, this amount of energy would mainly come from the area of
the fireball surface around the line of sight, i.e.,
$\theta\approx 0$ (where $\theta$ is the angle to the line of
sight). In other words, energy arising from the rest frame
high-energy portion (photons with $E_0>E_{0,p}$ when emitted)
emitted from the area with $\theta\approx 0$ would shift into the
channel concerned, and thus the peak time $t_p$ of the
corresponding light curves would keep almost unchanged (see Fig.
5). When $E_{p}$ increases, the contributions to the light curves
from the low-energy portion become larger since the corresponding
photons begin to shift into the channel concerned. At the same
time, some high energy photons emitted from the area corresponding
to a larger $\theta$ fall into the channel as well, and due to the
well-known time delay their contributions to the observed light
curves cause a larger $t_p$ (see Fig. 5). It is the difference of
$t_p$ of two different channels that leads to a lag (see Fig. 1).
When $E_{0,p}$ is fixed, the situation is quite different. In this
case, the enhancement of $E_{p}$ comes from the increasing of
$\Gamma$. A large $E_{p}$ corresponds to a large $\Gamma$ which
makes the light curve narrower (i.e., $FWHM \propto \Gamma^{-2}$,
see Qin et al. 2004; Lu et al. 2005a ). As a result, $t_p$ sharply
decreases with $E_{p}$ (see the upper panels of Fig. 4 and the
right panel of Fig. 5). Note that, since the influence of the
Lorentz factor is very significant (as $FWHM \propto
\Gamma^{-2}$), the difference of $t_p$ of two different channels
is hardly detectable in the right panel of Fig. 5.

Figs. 6 and 7 show that the lag increases with the low energy
spectral index $\alpha_0$ of the rest frame Band spectrum and it
decreases with the high energy spectral index $\beta_0$ of the
spectrum. The relationships also follow
$Lag_{G}<Lag_{13}<Lag_{14}$ in both cases. Compared with that
associated with the low energy spectral index, the variation of
the lag with respect to the increase of $\beta_0$ is relatively
mild, where $Lag_{13}$ and $Lag_{G}$ are almost unchanged (see
Fig. 7).

\begin{figure}
\centering
\includegraphics[width=5in,angle=0]{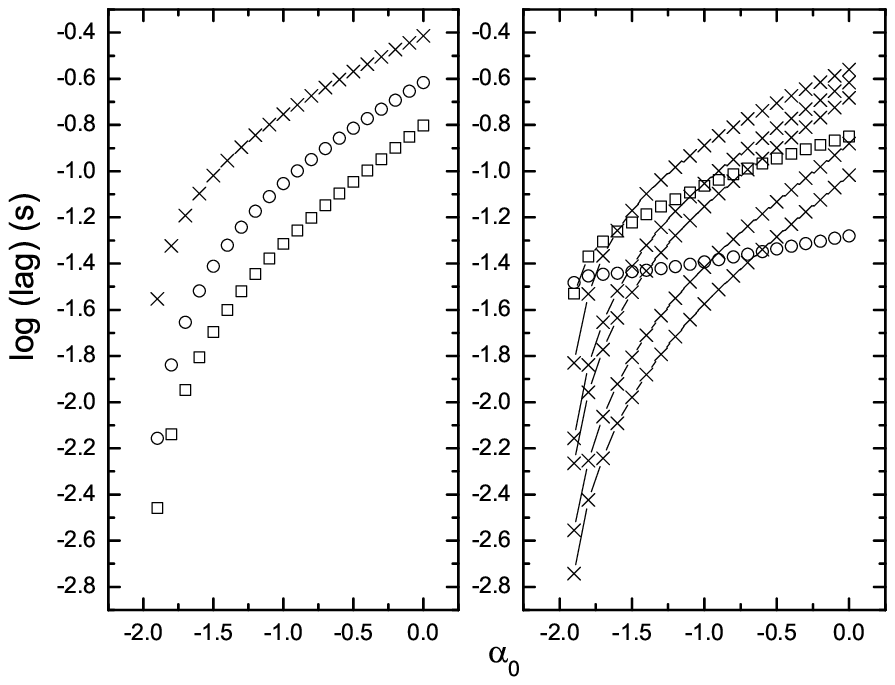} \caption{--Relationships between $Lag_{13}$, $Lag_{14}$,
$Lag_{G}$ and $\alpha_0$ for $E_{0,p}$=0.75 keV (the left panel),
and that between $Lag_{13}$ and $\alpha_0$ for different $E_{0,p}$
(the right panel). Here we take $\beta_{0}=-2.25$ and $\Gamma=200$.
In the right panel, the open circle and the open square represent
$E_{0,p}=0.1 keV, 0.15 keV$, respectively, crosses from the top to
the bottom represent $E_{0,p}=0.35 keV, 0.75 keV, 1 keV, 2 keV, 3
keV$, respectively. Other parameters and symbols are the same as
those adopted in Fig. 1.}
\end{figure}

\begin{figure}
\centering
\includegraphics[width=5in,angle=0]{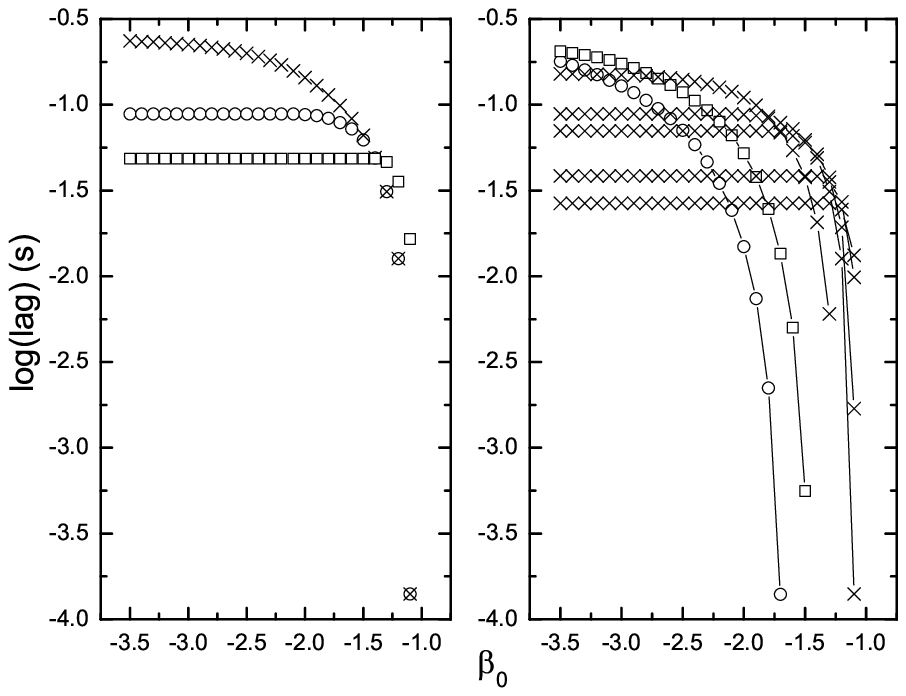} \caption{--Relationships between $Lag_{13}$, $Lag_{14}$,
$Lag_{G}$ and $\beta_0$ for $E_{0,p}$=0.75 keV (the left panel), and
that between $Lag_{13}$ and $\beta_0$ for different $E_{0,p}$ (the
right panel). Here we take $\alpha_{0}=-1$ and $\Gamma=200$. Other
parameters and symbols are the same as those adopted in Fig. 6.}
\end{figure}

To investigate what causes the relationships shown in Figs. 6 and
7, once more we divide the light curves of equation (1) into two
parts, as done in Fig. 4, here different $\alpha_0$ and $\beta_0$
are adopted (see Figs. 8 and 9).

\begin{figure}
\centering
\includegraphics[width=5in,angle=0]{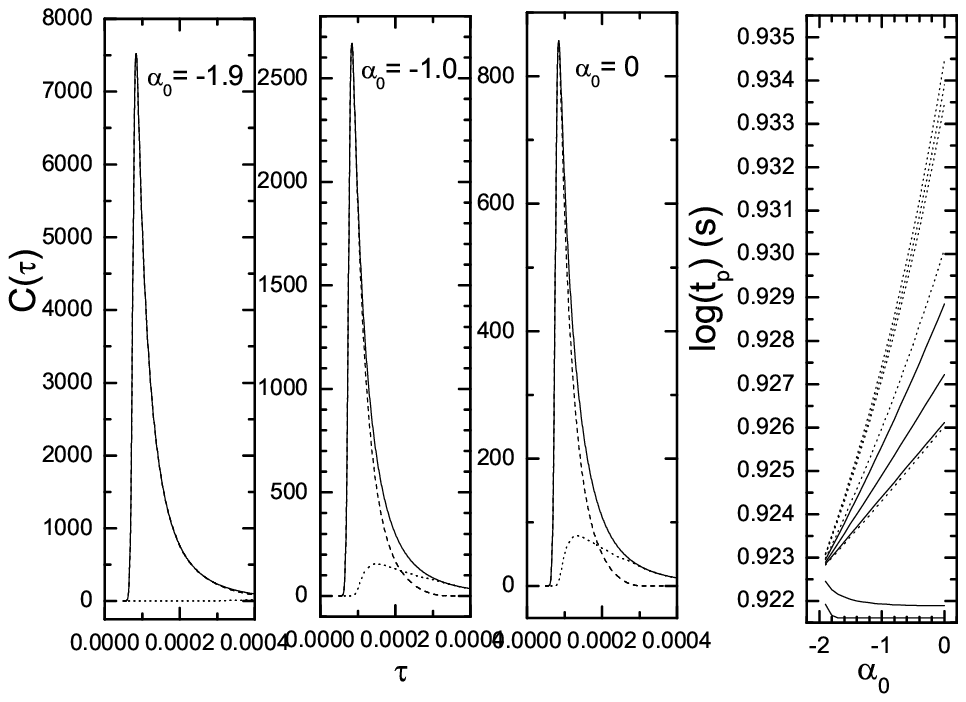} \caption{--Plots of the light
curves of BATSE channel 3 contributed from different portions of the
rest frame Band function spectrum for $E_{p}=250 keV$ (the first
three panels) and the relationship between the peak time of the
light curves of BATSE channel 3 (the solid lines) and 1 (the dot
lines) and $\alpha_0$ for different $E_{p}$ (the last panel). Here
we take $\beta_{0}=-2.25$ and $\Gamma=200$. In the first three
panels, the dot, dash and solid lines represent the light curves
contributed from the high-energy portion, the low-energy portion and
the whole energy range of the adopted spectra, respectively.  The
solid lines (or the dot lines) from the bottom to the top in the
last panel stand for $E_{p}=50, 100, 250, 300, 400$ keV,
respectively. The Other parameters are the same as those adopted in
Fig. 5.}
\end{figure}

\begin{figure}
\centering
\includegraphics[width=5in,angle=0]{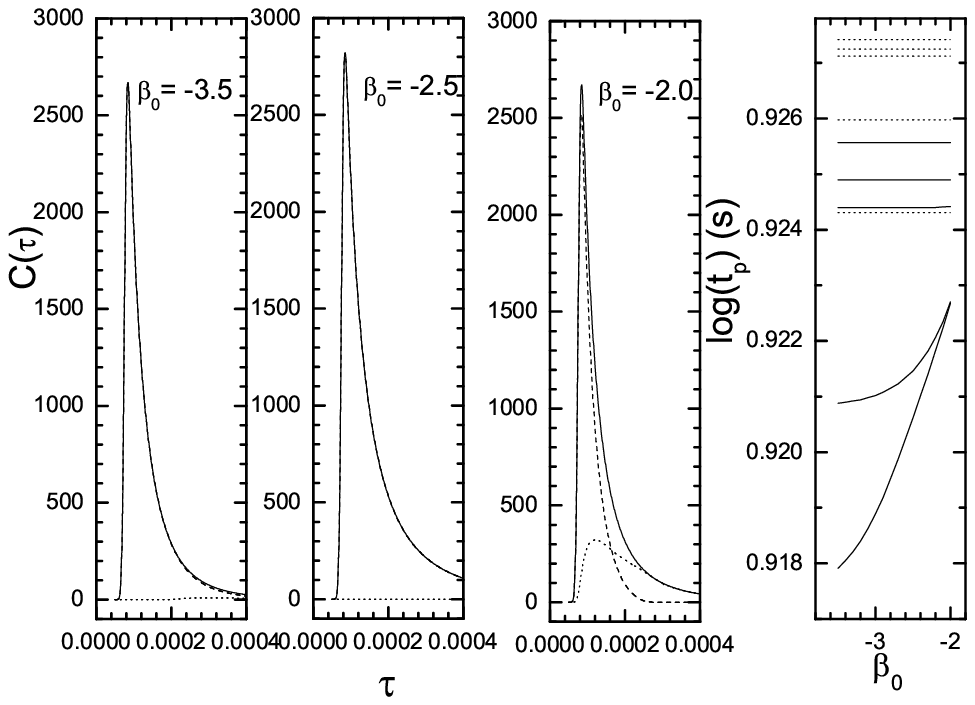} \caption{--Plots of the light
curves of BATSE channel 3 contributed from different portions of the
rest frame Band function spectrum for $E_{p}=250 keV$ (the first
three panels) and the relationship between the peak time of the
light curves of the light curves of BATSE channel 3 (the solid
lines) and 1 (the dot lines) and $\beta_0$ for different $E_{p}$
(the last panel). Here we take $\alpha_{0}=-1$. Other parameters and
symbols are the same as those adopted in Fig. 8.}
\end{figure}

Fig. 8 shows that, for BATSE channel 3 and in the case of $E_{p}=250
keV$, when $\beta_0$ is fixed, the contribution from the rest frame
low-energy portion to the light curve decreases slightly with the
increasing of $\alpha_0$, whereas the contribution from the rest
frame high-energy portion slightly increases with the increasing of
$\alpha_0$. The peak time $t_p$ of the first channel light curve of
BATSE rises more rapidly than that of the third channel (see Fig.
8). Note that, to shift into the channel observed, photons of the
rest frame high-energy portion must be those emitted from the area
of larger $\theta$, and it must be this that leads to the increasing
of $t_p$ with the increase of $\alpha_0$. The difference of $t_p$ of
the two channel light curves causes the lag (see Fig. 6).

From Fig. 9 we find a slightly different situation. For BATSE
channel 3 and in the case of $E_{p}=250 keV$, when $\alpha_0$ is
fixed, the contribution from the rest frame high-energy portion to
the light curve is significant only when $\beta_0$ is close to $-2$.
The peak time $t_p$ of the first channel light curve is almost
unchanged with the increase of $\beta_0$. Only when $E_{p}$ is low
enough, the contribution from the rest frame high-energy portion to
the light curve could become significant. In this situation, the
peak time $t_p$ of the third channel would rise with the increase of
$\beta_0$ (see Fig. 9). This is because that when $E_{p}$ is low
photons of the rest frame high-energy portion emitted from the area
of larger $\theta$ would shift into the channel observed, and then
the index $\beta_0$ would play a role.

\subsection{Spectral lag in the case of a rest frame spectrum varying with time}

It is common that the spectrum parameters of many GRBs are observed
to vary with time (see Preece et al. 2000). We are curious how the
lag's dependence on $\Gamma$ and $\Delta t_{\theta,FWHM}$ would be
if the rest frame spectrum develops with time. As most of GRBs show
``hard-to-soft" evolutionary feature (see e.g., Ford et al. 1995;
Fishman et al. 1995; Band, David L. 1997; Preece et al. 1998;), here
we consider a simple case where the rest frame spectrum is a Band
function with its indices and the peak energy $E_{0,p}$ decreasing
with time. Like Qin et al. (2005b), we assume a simple evolution of
indices $\alpha_0$ and $\beta_0$ and peak energy $E_{0,p}$ which
follow
$\alpha_0=-0.5-k(\tau_{\theta}-\tau_{\theta,1})/(\tau_{\theta,2}-\tau_{\theta,1})$,
$\beta_0=-2-k(\tau_{\theta}-\tau_{\theta,1})/(\tau_{\theta,2}-\tau_{\theta,1})$
and
$log\nu_{0p}=0.1-k(\tau_{\theta}-\tau_{\theta,1})/(\tau_{\theta,2}-\tau_{\theta,1})
$ $(keV h^{-1})$ for
$\tau_{\theta,1}\leq\tau_{\theta}\leq\tau_{\theta,2}$. For
$\tau_{\theta}<\tau_{\theta,1}$, $\alpha_0=-0.5$, $\beta_0=-2$ and
$log\nu_{0p}=0.1$ $(keV h^{-1})$, while for
$\tau_{\theta}>\tau_{\theta,2}$, $\alpha_0=-0.5-k$, $\beta_0=-2-k$
and $log\nu_{0p}=0.1-k$ $(keV h^{-1})$. We take k=0.1, 0.5
respectively (they correspond to different rates of decreasing) to
investigate the lag's dependence.

Let us employ local Gaussian pulse (2) to study this issue. We
adopt $\tau_{\theta,1}=9\sigma+\tau_{\theta,min}$ and
$\tau_{\theta,2}=11\sigma+\tau_{\theta,min}$ and
$\tau_{\theta,0}=10\sigma+\tau_{\theta,min}$ and
$\tau_{\theta,min}=0$.

\begin{figure}
\centering
\includegraphics[width=5in,angle=0]{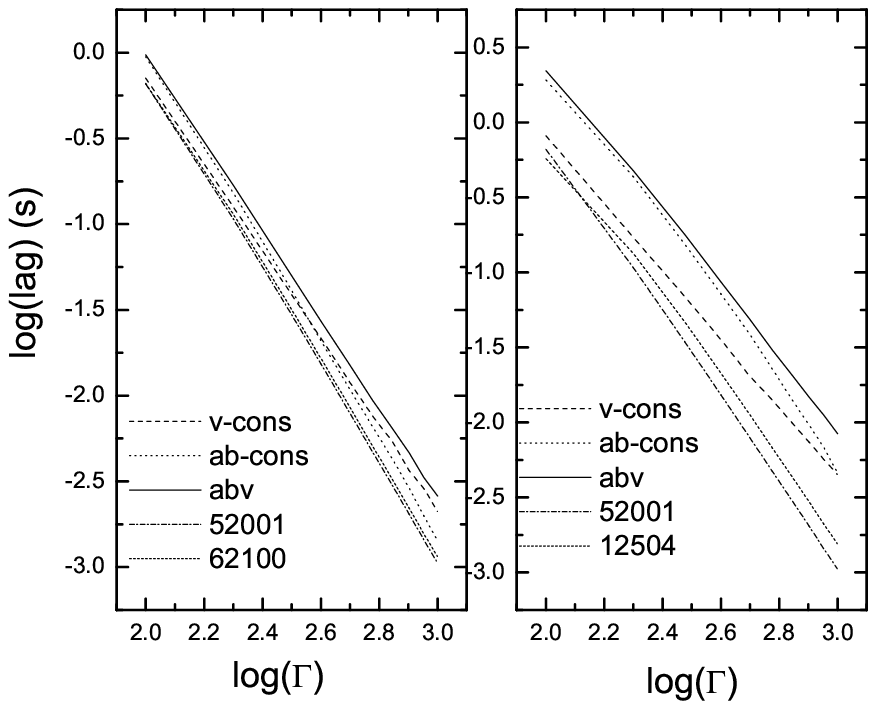} \caption{--Relationship between $Lag_{13}$ and
$\Gamma$ associated with the situation when the rest frame spectrum
varying with time in the case of $k=0.1$ (the left panel) and
$k=0.5$ (the right panel). For the sake of comparison, the
relationship deduced from the constant rest frame spectrum with the
corresponding upper and lower limits of the indexes are also
presented. The implications of the lines are as follows: v-cons
($\alpha_0$ and $\beta_0$ varying with time, $log\nu_{0p}=0.1$ $keV
h^{-1}$); ab-cons ($\alpha_0=-0.5$ and $\beta_0=-2.0$, $\nu_{0p}$
varying with time); abv ($\alpha_0$, $\beta_0$ and $\nu_{0p}$
varying with time); 52001 ($\alpha_0=-0.5$, $\beta_0=-2.0$ and
$log\nu_{0p}=0.1$ $keV h^{-1}$); 62100 ($\alpha_0=-0.6$,
$\beta_0=-2.1$ and $log\nu_{0p}=0.0$ $keV h^{-1}$); 12504
($\alpha_0=-1.0$, $\beta_0=-2.5$ and $log\nu_{0p}=-0.4$ $keV
h^{-1}$). Here we take $\Delta t_{\theta,FWHM}=10^{5} s$. Other
parameters are the same as those adopted in Fig. 1.}
\end{figure}

\begin{figure}
\centering
\includegraphics[width=5in,angle=0]{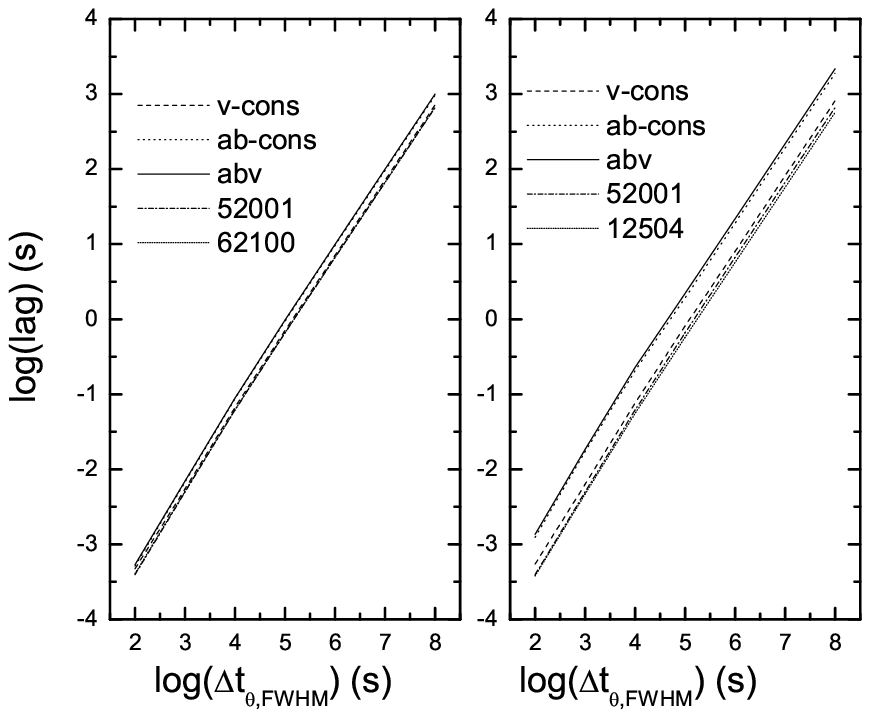} \caption{--Relationship
between $Lag_{13}$ and the local pulse width $\Delta
t_{\theta,FWHM}$ associated with the situation when the rest frame
spectrum varying with time in the case of $k=0.1$ (the left panel)
and $k=0.5$ (the right panel). For the sake of comparison, the
relationship deduced from the constant rest frame spectrum with the
corresponding upper and lower limits of the indexes are also
presented. Implications of the lines are the same as those in Fig.
10. Here we take $\Gamma=100$. Other parameters are the same as
those adopted in Fig. 1.}
\end{figure}

Figs. 10 and 11 show the dependence of lag on the Lorentz factor and
the local pulse width in the case when parameters of the rest frame
spectrum vary with time. We find that, when the spectral parameters
decrease with time, the lags are larger than that associated with
the case when the spectral parameters are fixed. The larger the
decreasing speed (say, k=0.5), the more obvious the effect.

\begin{figure}
\centering
\includegraphics[width=5in,angle=0]{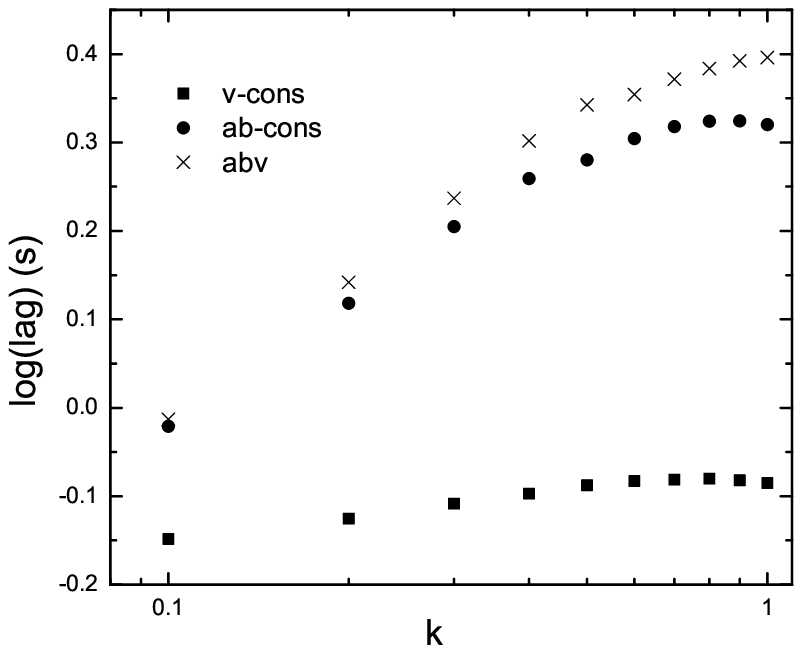} \caption{--Relationship
between $Lag_{13}$ and the varying speed $k$ of the parameters of
the rest frame spectrum. The implications of the lines are as
follows: v-cons ($\alpha_0$ and $\beta_0$ varying with time,
$logE_{0p}=0.1$ $keVh^{-1}$); ab-cons ($\alpha_0=-0.5$ and
$\beta_0=-2$, $E_{0p}$ varying with time); abv ($\alpha_0$,
$\beta_0$ and $\nu_{0p}$ varying with time). Here we take
$\Gamma=200$ and $\Delta t_{\theta,FWHM}=10^{5} s$. Other parameters
are the same as those adopted in Fig. 1.}
\end{figure}

Kocevski et al. (2003a) argued that the observed lag is the direct
result of spectral evolution and found the observed lag increases
with the peak energy's decay rate. Motivating by this, we take
$\alpha_0=-0.5$, $\beta_0=-2$, and allow $E_{0p}$ varying with time.
Illustrated in Fig. 12 is the relationship between the $Lag_{13}$
and the rest frame peak energy decaying rate, $k$, which is in
agreement with the observed result found by Kocevski et al. (2003a).
The dependence of lag on the varying speed $k$ in the case of the
rest frame spectrum parameters $\alpha_0$ and $\beta_0$ varying wit
time is also shown in Fig. 12. We find it common that the lag
increases with the increasing of $k$. The increasing of lag in the
case when the rest frame peak energy decreasing with time is much
more obvious than that in the case when indexes $\alpha_0$ and
$\beta_0$ decreasing with time.


\subsection{Lag's dependence on energy }

In a previous study, Wu \& Fenimore (2000) detected no larger lags
in lower energy bands. However, we find from the above analysis
that the following relation holds in almost all the cases
concerned: $Lag_{G}<Lag_{13}<Lag_{14}$. Motivated by this, we
investigate in the following in much detailed the time lag's
dependence on energy. We adopt a typical Band function spectrum
with $\alpha_0=-1$, $\beta_0=-2.25$ and $E_{0,p}=0.75 keV$ as the
rest frame radiation form to study this issue.

There are three instruments in the Swift telescope payload, and two
of them could detect high energy photons with the XRT ranging from
0.2 keV to 10 keV and the BAT ranging from 15 keV to 150 keV
(http:$//heasarc.gsfc.nasa.gov/docs/swift/about_swift/$). We
consider a wide band ranging from 0.2 keV to 8000 keV which covers
both the Swift and BATSE bands. This band is divided into the
following uniformly ranging channels (where $E_2=2E_1$): [$E_1$,
$E_2$]=[0.2, 0.4] keV, [0.5, 1] keV, [1, 2] keV, [2, 4] keV, [5, 10]
keV, [10, 20] keV, [20, 40] keV, [50, 100] keV, [100, 200] keV,
[200, 400] keV, [500, 1000] keV, [1000, 2000] keV, [2000, 4000] keV
and [4000, 8000] keV.

\begin{figure}
\centering
\includegraphics[width=5in,angle=0]{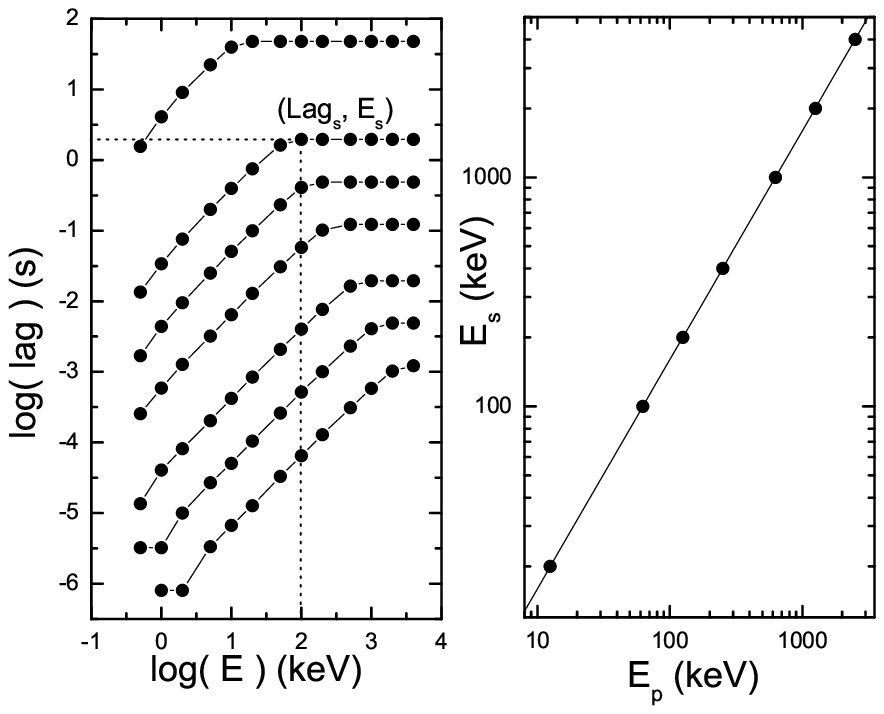} \caption{--Relationship between the
lag and the lower energy limit $E$ of the corresponding high energy
channel for different $\Gamma$ (the left panel) and that between
$E_s$ and $E_p$ (the right panel), where the relationship
$E_p=1.67\Gamma E_{0,p}$ is adopted and we take $\alpha_{0}=-1$,
$\beta_{0}=-2.25$ and $E_{0,p}=0.75 keV$. Solid lines from the top
to the bottom in the left panel stand for $\Gamma$=10, 50, 100, 200,
500, 1000, 2000, respectively. The solid line in the right panel is
the linear fit one of the data. Other parameters are the same as
those adopted in Fig. 1.}
\end{figure}

We measure the lags between the first channel and any of the other
channels. The relationship between the lag and the corresponding
energy $E$ (here $E$ denotes the lower energy limit of the
corresponding high-energy channel) is presented in Fig. 13. It shows
that, for any given value of $\Gamma$, the lags increase with energy
following the law of $lag \propto E$ and then saturate at a certain
energy (we call $E_{s}$, where subscript ``s" represents the word
``saturate"). This might explain why Wu \& Fenimore (2000) found
that GRBs do not show larger lag at lower energy. We notice that
$E_{s}$ takes place behind the corresponding peak energy $E_{p}$
(where the relation $E_p = 1.67 \Gamma E_{0,p}$ is adopted). For
different $\Gamma$, $E_{s}$ increases with $E_{p}$ following the law
of $E_{s}\propto E_p$ (see the right panel of Fig. 13), and the
relation $E_{s}\propto E_p$ holds for all $E_{0,p}$.

As shown in $\S2.2$, when $ E > E_{p}$, the contributions to the
corresponding light curve largely come from the high-energy
portion of the rest frame spectrum, which makes the peak time of
the light curve less change, and thus the lag would saturate. The
contrary is true. When $E \ll E_{p}$, the contributions come only
from the low-energy portion of the rest frame spectrum, emitted
from the whole fireball surface, i.e., $0\leq\theta \leq \pi/2$.
Thus, there would be almost no lag between the light curves of
such two energy channels (see the two filled circles in the bottom
of the left panel of Fig 13).

\subsection{Lag's dependence on the opening angle of uniform jets}

According to the relativistic fireball model, emissions from a
spherically expanding shell and a jet would be rather similar to
each other as long as we are along the jet's axis and the Lorentz
factor $\Gamma$ of the relativistic motion satisfies $\Gamma^{-1}<
\theta_{j}$, because the matter does not have enough time (in its
own rest frame) to expand sideways at such a situation (Piran
1995, 1999). Let us study the dependence of lag on the radiated
area confined by the opening angle of $\theta_{max}=p/\Gamma$,
where $p$ is a parameter describing the width of the opening angle
of uniform jets. Shown in Fig. 14 is the relationship between
$Lag_{13}$ and $p$.

\begin{figure}
\centering
\includegraphics[width=5in,angle=0]{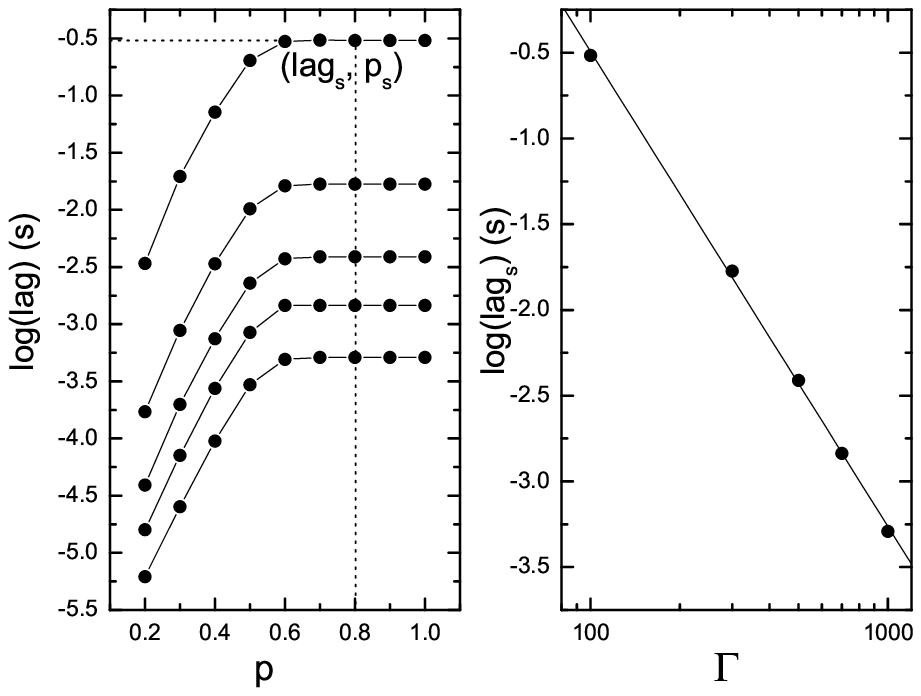} \caption{--Relationship between $Lag_{13}$ and
$p$ for different $\Gamma$ (the left panel), and that between
$lag_s$ and $\Gamma$ (the right panel), where we adopt a typical
Band function spectrum with $\alpha_0=-1$, $\beta_0=-2.25$ and
$E_{0,p}=0.75 keV$ as the rest frame radiation form. In left
panel, solid lines from the top to the bottom stand for
$\Gamma$=100, 300, 500, 700, 1000, respectively. Other parameters
are the same as those adopted in Fig. 1. }
\end{figure}

We find from Fig. 14 that $Lag_{13}$ increases with the opening
angle within $\theta_{max}<0.6\Gamma^{-1}$ following the law of
$Lag_{13}\propto p^{5.9}$, and beyond this range, it slightly
increases with the angle, and finally saturates at $p=0.8$ for any
$\Gamma$. The saturate lag (denoted by $Lag_s$) decreases with the
increasing of $\Gamma$ following the law of $Lag_s \propto
\Gamma^{-2.8}$ (see the right panel of Fig. 14). This is an
outcome of the beaming effect. It suggests that there is a limited
value of the lag when different opening angles of uniform jets are
concerned, which is the same as that in the case of spherical
fireballs.

\subsection{Lag's dependence on the radius of fireball }

Shen et al. (2005) found that the lag is independent of the radius
of fireballs based on their model (see Fig. 8 in their paper). We
perform the analysis with their formula when raising a higher
precision of calculation (the result is the same when we adopt
equation [1]). Here we fix the local pulse width by adopting
$\Delta t_{\theta}=10 s$. The result is presented in Fig. 15.

\begin{figure}
\centering
\includegraphics[width=5in,angle=0]{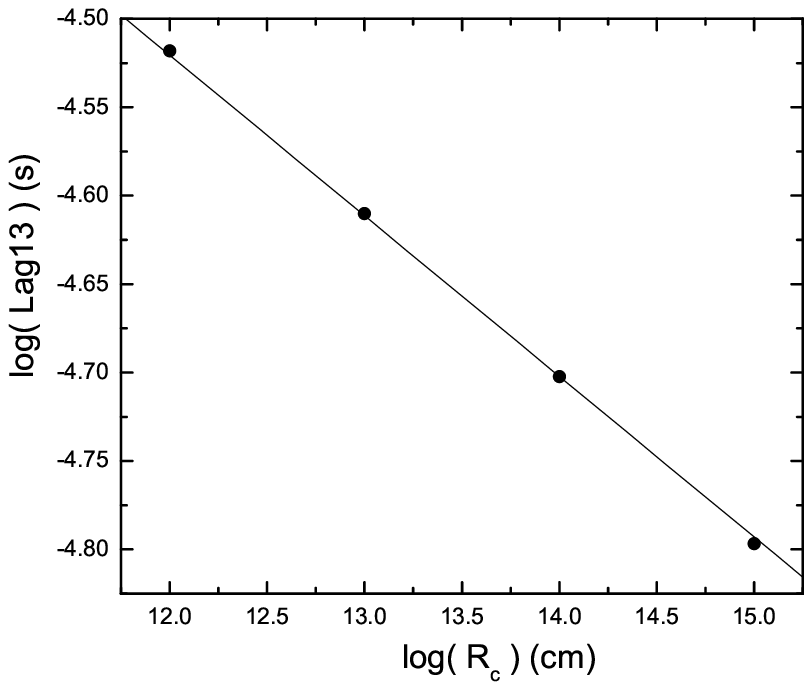} \caption{--Relationship between $Lag_{13}$ and the radius of
the fireball. Here we take $\Gamma=100$, $E_{0,p}=0.75 keV$ and
$\Delta t_{\theta,FWHM}=10 s$. The solid line is a linear fit to the
data. Other parameters are the same as those adopted in Fig. 1. }
\end{figure}

Fig. 15 shows that $Lag_{13}$ decreases with the radius of fireballs
following the law of $Lag_{13}\propto R_{c}^{-0.1}$, when the local
pulse width is fixed. The lag depends on the radius of fireballs.
Comparing our result with that obtained in Shen et al. (2005) we
find that the conclusions are entirely different. We suspect that
three factors might be responsible to this difference. The first is
the precision of calculation. The second is the range concerned. Our
calculation spans about three magnitudes of the fireball radius
while Shen et al. (2005) considered only two magnitudes. The third
is that the dependence of $Lag_{13}$ on the radius is weak: the lag
would be only about two times smaller when the radius becomes three
magnitudes larger. We believe that the third is the key factor
accounting for the difference.

The conclusion of Shen et al. (2005) that the lag is independent
of the radius of fireballs is a puzzle since a large radius seems
to cause a larger distance between the area with $\theta=0$ and
that with $\theta>0$ on the fireball surface and this should lead
to a larger lag. Our conclusion that the lag decrease slightly
with the increasing of the radius makes the puzzle worse. The
mechanism accounting for this is currently unclear. We suspect
that this dependence might rely on what kind of photons dominate
the peaks of the lower and higher energy channels respectively and
where they are emitted from. In creating the dependence, other
parameters such as the rest frame peak energy and the Lorentz
factor might be at work. Therefore, this might not be answerable
if only a simple mechanism is concerned (to find a answer to this,
a detailed investigation should be made, which is beyond the scope
of this paper).

\section{Different characteristics shown in the two cases of fixing the intrinsic pulse width and the local pulse width}

Shen et al. (2005) explored the lag's dependence on the Lorentz
factor and found $Lag \propto \Gamma^{-1}$ when fixing the
intrinsic pulse width (the co-moving pulse width) and the fireball
radius. However, in this paper we find $Lag \propto \Gamma^{-2}$
instead, when fixing the local pulse width and the fireball radius
based on both models of Shen et al. (2005) and Qin et al. (2004).
According to the theory of special relativity, the two results are
identical, as the local timescale is $\Gamma$ times of the
corresponding intrinsic timescale. Some interesting questions
arise accordingly. When the fireball radius is fixed (as
considered in Shen et al. 2005 and in this paper), would the two
approaches always lead to identical results? Could they reveal the
same property of the curvature effect? In which case should which
one be preferred?

As pointed out and illustrated in Qin et al. (2004), the profiles
of light curves are not affected by the Lorentz factor when fixing
the relative local pulse width ($c\Delta t_{\theta}/R_{c}$, or
$\Delta \tau_{\theta}$ as used in Qin et al. 2004). When the
fireball radius is fixed, a fixed relative local pulse width would
correspond to a fixed local pulse width (that is, a certain
$\Delta \tau_{\theta}$ leads to a certain $\Delta t_{\theta}$ in
this situation, where $R_{c}$ is fixed). Thus, in this situation,
fixing the local pulse width would not lead to different profiles
of light curves when different Lorentz factors are considered.
However, the feature of the profile will change when fixing only
the intrinsic pulse width. This is because that for a ceratin
intrinsic pulse width, the corresponding local pulse width would
take different values when different Lorentz factors are
considered (recall that the local timescale is $\Gamma$ times of
the corresponding intrinsic timescale), and this would lead to
different relative local pulse width since $R_{c}$ is fixed. The
corresponding profiles of light curves would change since they are
sensitive to the relative local pulse width (see Qin et al. 2004).

Fig. 16 illustrates the different influences of the Lorentz factor
on the profiles of the light curves calculated with equation (5)
in Shen et al. (2005) in the two cases of fixing the local pulse
width and the intrinsic pulse width.

\begin{figure}
\centering
\includegraphics[width=5in,angle=0]{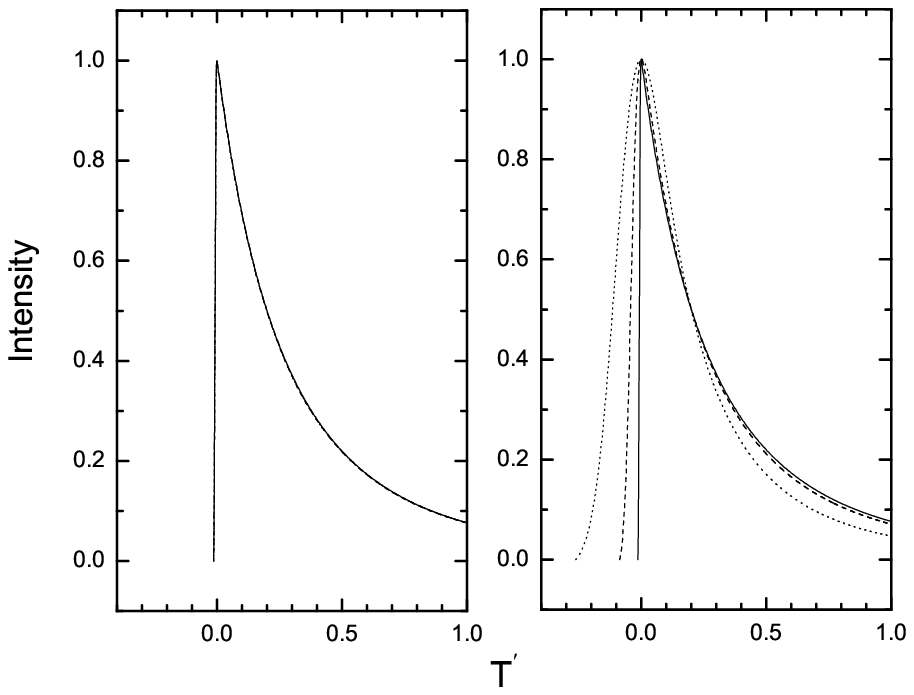} \caption{--Plots of the light
curves of BATSE channel 3 resulting from symmetric Gaussian pulses
when fixing the local pulse width $td=100 s$ (the left panel) and
the intrinsic pulse width $td^{\prime}=10 s$ (the right panel) for
different $\Gamma$. Here we take $\alpha=-1$, $\beta=-2.25$,
$R_c=5\times 10^{14} cm$ and $E_{p}=350 keV$ (where the relationship
$E_p=1.67\Gamma E_{0,p}$ is adopted). The solid, dash and dot lines
stand for $\Gamma=10$, $100$, $1000$ respectively (the three lines
in the left panel overlap each other). Note that, for the sake of
comparison, we normalize the light curves and re-scale the variable
so that the peak count rate is located at $T^{\prime}=0$ and the
FWHM of the decay portion is located at $T^{\prime}=0.2$. }
\end{figure}

Kocevski et al. (2003b) and Ryde et al. (2003) found that there is a
linear relationship between the rise width, $FWHM_{r}$ and the full
width, FWHM, of gamma-ray pulses. Lu et al. (2005a) explained the
relationship based on the formula presented in Qin et al. (2004),
and found that there exists a dead line in the $FWHM_{r} - FWHM$
plane when taking the relative local pulse width $\Delta
\tau_{\theta,FWHM}\geq 1$. Here we repeat the same work as
illustrated in Fig. 2 of Lu et al. (2005a) based on the Shen et al.
(2005) formula. As shown in Fig. 17, we find that there indeed
exists the dead line predicted before in the $FWHM_{r} - FWHM$ plane
when fixing the local pulse width (here we take $td\geq1000 s$ and
$R_c=3\times 10^{13} cm$ which corresponds to $\Delta
\tau_{\theta,FWHM}\geq 1$). However, when fixing the intrinsic pulse
width, one cannot find a dead line in the plane (see Fig. 17). This
indicates that the features in the $FWHM_{r} - FWHM$ plane are
different in the case of fixing the local pulse width and in that
fixing the intrinsic pulse width. One could obtain the local pulse
width td by measuring the location of the corresponding data (to
observe in what lines they belong to) from the left panel of Fig. 17
as long as $\Delta \tau_{\theta,FWHM}< 1$. But one could not measure
the intrinsic pulse width from the right panel of the figure. This
indicates that the former is superior to the latter when one
attempts to make use of the $FWHM_{r} - FWHM$ plane.

\begin{figure}
\centering
\includegraphics[width=5in,angle=0]{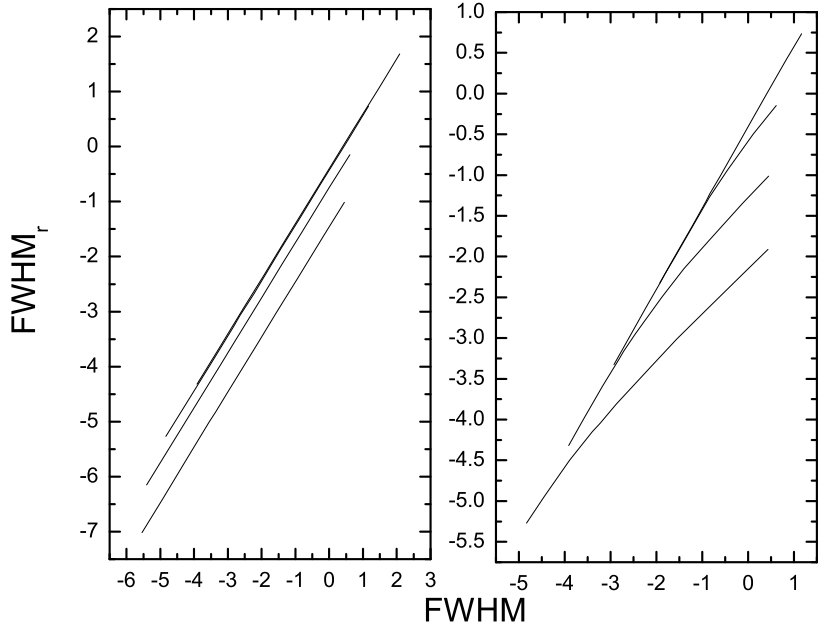} \caption{Relationships between $FWHM_r$ and $FWHM$ for
the light curves of BATSE channel 3 resulting from symmetric
Gaussian pulses associated with the case of fixing the local pulse
width (the left panel) and the case of fixing the intrinsic pulse
width (the right panel). In the left panel, solid lines from the
bottom to the top represent $td=10 s$, $100 s$, $1000 s$, $10000 s$,
respectively (note that the last two solid lines overlap each
other), and in the right panel, the corresponding lines stand for
$td^{\prime}=0.1 s$, $1 s$, $10 s$, $100 s$, respectively. Here we
take $R_c=3\times 10^{13} cm$ and other parameters are the same as
those adopted in Fig. 16.}
\end{figure}

\section{Discussion and conclusions  }

Discussed in Shen et al. (2005) and the above analysis, only
intrinsic or local Gaussian pulses are involved. One might ask if
the conclusions hold only in the selected pulse form. Thus, we
replace the local Gaussian pulse with other forms of local pulses
when applying equation (1) and the corresponding formula of Shen et
al. (2005). Local pulses (55), (81), (82), (83), (85), (86) and (87)
presented in Qin et al. (2004) are studied. We find that the lags
resulting from a local pulse without a decaying phase are too small
to be noticed. For example, $lag13_{t} < 10^{-6} s$ when
$\Gamma>100$ and $R_c=3\times10^{15} cm$. This is in well agreement
with what discovered by Shen et al. (2005). Now it is clear that, in
the mechanism of the curvature effect, it is the decaying phase of a
local pulse that contributes to the observed lags.

A recent study revealed that (see Lu et al. 2005b), the rise phase
of light curves will always be dominated by the emission from
$\theta \simeq 0$ and within the rise phase of the local pulse. It
is illustrated in Fig. 18, where a local pulse without a decaying
phase is considered. This indicates that there would be no peak
lags for the light curves of different energy channels associated
with those local pulses without a decaying phase.
\begin{figure}
\centering
\includegraphics[width=5in,angle=0]{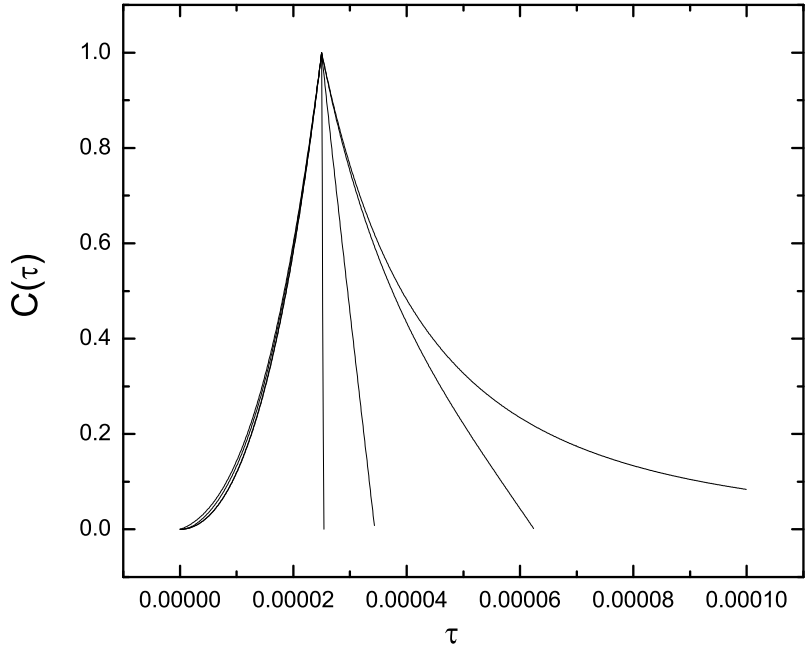} \caption{Plots of the
normalized light curves of BATSE channel 3 arising from a local
linear rise pulse with $\Delta \tau_{\theta,FWHM}=1$ (a local pulse
without a decaying phase) emitted from different areas of the
fireball surface. Here we take $\Gamma=200$ and adopt a Band
function rest frame radiation form with $\alpha_0=-1$,
$\beta_0=-2.25$ and $\nu_{0p}=0.75 keV h^{-1}$. The solid lines from
the right to the left stand for $\theta_{max}$=$\pi/2$, $1/\Gamma$,
$0.5/\Gamma$ and $0.1/\Gamma$, respectively (note that the rising
phase as well as the peak of the counts rates for different light
curves overlap each other). Other parameters are the same as those
adopted in Fig. 1.}
\end{figure}

Except the Band function spectrum, the Comptonized spectrum and the
broken power law spectrum were employed to fit the spectral data in
Preece et al. (2000). To make sure if the dependence of lags
discussed above would be altered when other rest frame spectra are
considered, we employ the Comptonized spectral model as the rest
frame radiation form to investigate the lag's dependence concerned
above. The Comptonized spectrum is written as $I_\nu \propto
\nu_{0,\theta}^{1+\alpha_{0,C}}exp(-\frac{\nu_{0,\theta}}{\nu_{0,C}})$,
where $\alpha_{0,C}$ and $\nu_{0,C}$ are constants. Here, typical
values, $\alpha_{0,C}=-0.6$ (Schaefer et al. 1994) and
$\nu_{0,C}=0.55 keVh^{-1}$ (see Qin 2002 Table 2) are adopted. Our
calculation shows that the characteristics of the lag's dependence
are the same as what revealed above, when the rest frame Band
function spectrum is replaced with the two other forms. The only
difference is the magnitude of the lags. For example, when the rest
frame Comptonized spectral model is adopted, we obtain a slightly
larger value of $Lag_{13}$ (about $70\%$ larger) when we redraw Fig.
2.

Of the sample of 41 source presented in Friedman et al. (2005), we
find that the range of $\theta_{j}$ is \{0.03107, 0.57072 \} rad,
and its average value is 0.15925 rad. It shows that $\theta_{j}
\gg \Gamma^{-1}$ if $\Gamma>100$. As discussed in \S 2.5, one
could find that for GRBs coming from a uniform jet with such
opening angles, if $\Gamma>100$ is adopted, the difference of lags
from that arising from a spherical surface would not be
distinguishable.

Several authors showed that the CCF lags between BATSE channels 1
and 3 tend to concentrate near $<200ms$ (see Norris et al. 1996,
2000; Chen et al. 2005). A recent measurement carried by Norris et
al. (2005) for BATSE wide pulses showed that the long lag range is
1.0 $< lag <$ 4.2 s. It is generally believed that the gamma-rays
arise from internal shocks at a distance of $R \sim 10^{13}-10^{15}$
cm from the initial source (Piran 2005). Within the generally
accepted ranges of $R_c = 10^{13}-10^{15}$ cm and $\Gamma>100$, the
observed lags could be accounted for by the curvature effect when
different values of parameters are adopted. For example, one gets
$Lag_{13} = 300$ ms from Fig. 1 when taking $\Gamma= 100$; $Lag_{13}
= 100 ms$ from Fig. 7 when taking $\alpha_0 = -1$, $\beta_0 = -2.25$
and $E_{0,p} =0.75 keV$. According to Fig. 2, $Lag_{13}\simeq
4.6\times10^{-7}\Delta t_{\theta,FWHM}$ s. One could obtain a large
lag, i.e., $Lag_{13} = 4.6$ s when taking $\Delta t_{\theta, FWHM} =
10^{7} s$. In this situation, the $FWHM$ of the corresponding light
curve determined by equation (1) is about 75 s when taking $\Gamma=
500$, which is very close to the observed data of the wide-pulse
bursts studied in Norris et al. (2005), i.,e., $1.0 < lag < 4.2$ s
and $6s<$ the width of channel 1 $<100 s$ (see Fig. 6 in that
paper). All these show that merely the curvature effect could
produce the observed lags. A larger lag requires a wider local
pulse.

We thus come to the following conclusions: a) lag $\propto
\Gamma^{-\epsilon}$ with $\epsilon>$ 2 within different energy
channels, and the relationship would different for different rest
frame peak energy $E_{0p}$; b) lag is proportional to the local
pulse width and the $FWHM$ of the observed light curves, and the
local pulse width plays a more important role in producing a large
lag when compared with other lag's dependent parameters; c) a large
lag requires a large $\alpha_0$ and a small $\beta_0$ as well as a
large $E_{0p}$; d) when the rest frame spectrum varies with time,
the lag would become larger; e) lag decreases with the increasing of
$ R_c$; f) lag $\propto$ E within a certain energy range for a given
$\Gamma$; g) lag is proportional to the opening angle of uniform
jets when $\theta_{max} < 0.6\Gamma^{-1}$.

We are very grateful to Dr. Rong-feng Shen for sending us his
valuable program and for his helpful suggestions. This work is
supported by the Special Funds for Major State Basic Research
Projects (``973'') and National Natural Science Foundation of
China (No. 10273019 and No. 10463001).

\label{lastpage}


\clearpage
\begin{thebibliography}{10}

\bibitem[Band et al. (1993)]{Ba93}  Band, D.,Matteson, J., Ford, L., Schaefer, B., Palmer, D.,
 Teegarden, B., Cline, T., Briggs, M., et al. 1993, ApJ, 413, 281
\bibitem[Band, David L.  (1997)]{Ba97} Band, David L. 1997, ApJ, 486, 928B

\bibitem[Chen et al. (2005)]{Ch05}Chen, Li, Lou, Yu-Qing, Wu, Mei, Qu, Jin-Lu, Jia, Shu-Mei, Yang, Xue-Juan. 2005, ApJ, 619, 983
\bibitem[Cheng et al. (1995)]{Ch95}Cheng L. X., Ma Y.Q., Cheng K. S., Lu T. \& Zhou Y. Y. 1995, A\&A, 300, 746

\bibitem[Fishman et al. (1995)]{Fi95} Fishman, G., Meegan, C. 1995, ARA\&A, 33, 415
\bibitem[Ford et al. (1995)]{Fo95} Ford, L. A., Band, D. L., Matteson, J. L., Briggs, M. S.,
 Pendleton, G. N., Preece, R. D., Paciesas, W. S., Teegarden, B. J., et al. 1995, ApJ, 439, 307F

\bibitem[Friedman et al. (2005)]{Fr05} Friedman, Andrew S. \& Bloom, Joshua S. 2005, ApJ,627, 1F

\bibitem[Ioka et al. (1999)]{Io01} Ioka K. \& Nakamura T. 2001, ApJ, L163

\bibitem[Kocevski et al. (2003)]{Ko03a}  Kocevski D. \& Liang E. 2003a, ApJ, 594, 385
\bibitem[Kocevski et al. (2003)]{Ko03b}  Kocevski D., Ryde F., \& Liang E., 2003b, ApJ, 596, 389
\bibitem[Lu et al. (2005)]{La05a}Lu, R. J., Qin, Y. P., \& Yi, T. F., 2005a, Chin. J. Astron. Astrophys., in press, astro-ph/0508599
\bibitem[Lu et al. (2005)]{La05b}Lu, R. J., Qin, Y. P., 2005b, MNRAS, Submitted, astro-ph/0508537

\bibitem[Norris et al. (1996)]{No96}  Norris, J. P., Nemiroff, R. J., Bonnell, J. T.,
 Scargle, J. D., Kouveliotou, C., Paciesas, W. S., Meegan, C. A., Fishman, G. J. 1996, ApJ, 459, 393
\bibitem[Norris et al. (2000)]{No00}  Norris, J. P., Marani G. F., \& Bonnell J. T. 2000, ApJ, 534, 248
\bibitem[Norris et al. (2001)]{No01}Norris J. P., Scargle, J. D. \& Bonnell, J. T. 2001, astro-ph/0105052
\bibitem[Norris et al. (2005)]{No05}  Norris, J. P., Bonnell, J. T., Kazanas, D., Scargle, J. D.,
 Hakkila, J., Giblin, T. W. 2005, ApJ, 627, 324N

\bibitem[Piran (1995)]{Pi95}Piran, T. 1995, in AIP Conf. Proc. 307, Gamma-Ray Bursts: Second Huntsville
Workshop, ed. G. J. Fishman, J. J. Brainerd, \& K. Hurley (New
York: AIP), 495
\bibitem[Piran (1999)]{Pi99}Piran, T. 1999, Phys. Rep., 314, 575
\bibitem[Piran (2005)]{Pi05} Piran, T. 2005, Rev. Mod. Phys. 76, 1143

\bibitem[Preece et al. (1998)]{Pr98}  Preece, R. D., Pendleton, G. N.,
Briggs, M. S.,Mallozzi, Robert S., Paciesas, William S., Band,
David L., Matteson, James L., Meegan, C. A. 1998, ApJ, 496, 849
\bibitem[Preece et al. (2000)]{Pr00}  Preece, R. D., Briggs, M. S.,
Mallozzi, R. S., Pendleton, G. N., Paciesas, W. S. and Band, D. L.
2000, ApJS, 126, 19

\bibitem[Qin (2002)]{Qi02}  Qin, Y.-P. 2002, A\&A, 396, 705
\bibitem[Qin (2003)]{Qi03}  Qin, Y.-P. 2003, A\&A, 407, 393
\bibitem[Qin et al. 2004]{Qi04}Qin, Y. P., Zhang Z. B., Zhang F. W. and Cui X. H. 2004, ApJ,
617, 439 (Paper II)
\bibitem[Qin \& Zhang (2005)]{Qi05a} Qin, Y.-P., \& Zhang, B.-B. 2005a, astro-ph/0504070
\bibitem[Qin et al. (2005)]{Qi05b} Qin, Y.-P., \& Dong, Y.-M., \& Lu R.-J., B.-B. Zhang, L.-W. Jia 2005b,
ApJ, 632, 1008

\bibitem[Ryde et al. (2003)]{Ry03}Ryde F., Borgonovo L., Larsson S. et al., 2003, A\&A, 411, L331

\bibitem[Schaefer et al. (1994)]{Sc94} Schaefer, Bradley E., Teegarden, Bonnard J., Fantasia, Stephan
F., Palmer, David, Cline, Thomas L., Matteson, James L., Band,
David L., Ford, Lyle A., et al. 1994, ApJS, 92, 285S
\bibitem[Schaefer et al. (2004)]{Sc04}  Schaefer, Bradley E. 2004, ApJ, 602, 306
\bibitem[Shen et al. (2005)]{Sh05}R.-F. Shen, L.-M. Song, Z. Li, 2005, MNRAS, 362, 59S (Paper I)
\bibitem[Salmonson (2000)]{Sa00}Salmonson J. D. 2000, ApJ, 544, L115

\bibitem[Wu et al. (2000)]{Wu00}Wu B. \& Fenimore E. 2000, ApJ, 535, L29

\end{thebibliography}
\end{document}